# Heart rate and its variability as an indicator of mental health in male prisoners

Christian Gold[1,2], Jörg Assmus[1,3]
[1] GAMUT, Uni Research Health, Uni Research, Bergen, Norway
[2] Grieg Academy Department of Music, University of Bergen, Norway
[3] Centre for Clinical Research, Haukeland University Hospital, Bergen, Norway

Correspondence
Christian Gold
GAMUT – The Grieg Academy Music Therapy Research Centre
Uni Research Health
Uni Research
Lars Hilles gt. 3
5015 Bergen
Norway
Phone +47-97501757
Email: christian.gold@uni.no

**Abstract**
Heart rate (HR) and its variability (HRV) has been proposed as a marker for depressive symptoms and other aspects of mental health. However, the real correlation between them is presently uncertain, as previous studies have generally been conducted on the basis of small samples. In a sample of 113 adult male prisoners, we analyzed correlations between five measures of HR/HRV and five psychological measures of mental health aspects (depression, state and trait anxiety, and social relationships). We used Nadaraya-Watson non-parametric regression in both directions and age-stratified Spearman correlation to detect possible relations. Despite strong correlations among HR/HRV measures and among psychological measures, correlations between HR/HRV and psychological measures were low and non-significant for the overall sample. However, we found an age dependency, suggesting some correlations in younger people (HR with STAI-State, $r = 0.39$; with HADS-Anxiety, $r = 0.52$; both $p < .005$). Overall, the general utility of HR/HRV as a marker for mental health across populations remains unclear. Future research should address age and other potential confounders more consistently.

## i. Introduction

In evidence-based mental health care, it is important, but not always straightforward, to choose the best measures to assess outcomes. Already selecting the right domain can be challenging – for example, one has to choose between ‘positive’ or ‘negative’ outcome domains (e.g., symptoms or functioning), and between proximal (direct) or distal (downstream) outcomes, all to reflect best what the intervention can do and what clients and providers need or request (1). In addition, choosing the right data source for a given domain is more than a technicality. A relatively recent challenge in this context is the choice between traditional psychological assessments (i.e., either self-report or assessor-based, often using questionnaires) versus newer neurophysiological measures. Neurophysiological assessments rely on technological tools for measuring some indicator of brain activity. If successful, such measures may provide a more objective basis for judging someone's mental health, as they are less susceptible to purposeful distortions and biases than traditional psychological assessments. In addition, they might also help to understand links between mental and somatic processes. However, in contrast to questionnaire-based methods, physiological indicators are usually 'found' rather than developed for the purpose. They are therefore reflective of a variety of factors, including many that may be unrelated to the domain of interest and that will complicate the use of these measures by acting as noise or confounding variables (2). For example, a recent study showed almost no correlation between encephalographic markers for depression/anxiety and psychological assessments of these domains (3). However, many recent studies have used or suggested physiological indicators of mental health as outcomes, either alongside or instead of psychological assessments (4-11).

Heart rate variability (HRV) is one such marker. As a potential biomarker, it is "important not so much for what it tells us about the state of the heart as much as it is important for what it tells us about the state of the brain" (12). HRV is influenced by various influences from the brain, including both the sympathetic as well as the parasympathetic nervous system. Simply put, sympathetic input increases the heart rate (in response to stressors, to facilitate 'fight or flight' behaviors), and parasympathetic input decreases it (to enable relaxation, to facilitate 'rest and digest' behaviors). High HRV might therefore indicate that both systems work well and are in balance; the individual is able to respond and adapt to stressful situations as well as to relax. HRV has therefore been suggested as a potential biomarker for numerous aspects of mental health: It could be a "marker of stress and health" (12), or of "stress and resilience" (p. 751). It has also been suggested to reflect the "link between emotional states and dispositions such as depression, anxiety, anger and hostility, alexithymia, and physical health" (p. 751) and even "post-traumatic stress disorder, and schizophrenia" (p. 750). Which of these it should reflect most remains however unclear. It may be a non-specific indicator of mental health rather than specific to certain disorders. Average heart rate (HR), an older and simpler measure than HRV, may also indicate aspects of mental (as well as physical) health (13, 14).

The potential link of HRV with depression has been studied most extensively. A recent meta-analysis examined the association between HRV and depression in people without cardiovascular disease (15). Based on four included studies (total n=200), a moderately strong correlation between HRV and depression severity was found ($r = -0.35$, $p < 0.001$). However, the included studies were small. A larger study with negative findings was excluded from the meta-analysis (16), partly because it was unpublished (17). Publication bias – the selective publication of studies with positive results – may be an issue (18). In addition, given that several measures of HRV exist, selective reporting of positive measures within studies may also be associated with overestimating the true association (18). It remains therefore unclear how well HRV really performs as a potential biomarker of depression. Furthermore, it has been suggested that associations between HRV and mental health may be non-linear (15, 17), but previous studies have relied on linear correlation analysis. Finally, correlations with other aspects of mental health have not been examined systematically. A recent meta-analysis suggested categorical differences in HRV between people with and without anxiety disorders, but did not examine correlations with the severity of anxiety symptoms (19). One study suggested that social relationships may also be related to HRV (20).

HRV may also be relevant in prisoners. Some prisoners have alleviated levels of depression and anxiety (21). Several studies have investigated possible links between antisocial behavior and HR (22) or between psychopathy and anxiety (23), but these links are not yet clearly understood.

The present study aimed at examining linear and non-linear associations between various aspects of mental health – standardised measures of depression, anxiety, and quality of social relationships – and different measures of HR/HRV, in male adult prisoners. Specifically, we hypothesized that HRV would be positively correlated to positive aspects of mental health (social relationships) and negatively to its negative aspects (depression and anxiety), and vice versa for HR.

## ii. Material and Methods

### 2.1 Participants

We used an existing sample of 113 adult male prisoners originally recruited for an intervention study (21) (trial registration: ISRCTN22518605). Ethical approval was granted by the Regional Committee for Medical and Health Research Ethics Western Norway (REK Vest). Prisoners have on average higher levels of mental health problems than the general population and are a high-risk group for mental disorders (24, 25). All inmates at Bjørgvin prison, Norway, who had sufficient command of the Norwegian language and consented to participate in the study were eligible. They were given verbal and written information about the project and were included if they decided to sign an informed consent statement. Participants were between 18 and 64 years of age (M = 31.38, see Table 1). Expected stay at the prison ranged widely from 4 to 840 days, with a median of 42 days. (This variable is reported purely for descriptive purposes and was not used in the analysis.) Levels of self-reported anxiety and depression were somewhat alleviated in the sample; 45% had clinically relevant levels of anxiety and 29% had clinically relevant depression (Table 1). These cutoffs were defined by previous research (26) and correspond to criteria for generalized anxiety disorder or major depressive episode, respectively.

### 2.2 Psychological measures

We used well-established psychological measures of anxiety, depression, and social relationships, including the State-Trait Anxiety Inventory (27, 28) with subdomains state anxiety (STAI-State) and trait anxiety (STAI-Trait); the Hospital Anxiety and Depression Scale (26, 29, 30) with subdomains anxiety (HADS-A) and depression (HADS-D); and the Social Relationships scale of the Quality of Life Enjoyment and Satisfaction Questionnaire, Q-LES-Q (31). Internal consistency of these measures in the present sample was high (ranging from 0.77 for HADS-D to 0.94 for STAI-State).

### 2.3 Physiological measures

HR and HRV was registered using the Actiheart System (32), a compact lightweight device that records physical activity, HR, and variability of cardiac inter-beat interval. The Actiheart clips onto a single ECG electrode (Cleartrode, Disposable Pregelled Electrodes, 150, Standard Silver) with a short ECG lead to another electrode that picks up the ECG signal. The electrode was placed on the upper chest. Resting (baseline) psychophysiological activity was registered for five minutes while the participants were seated in a comfortable chair. The inter-beat-interval data was manually cleaned. A start and an end point were selected, and the area was replaced with the previous 15 seconds. As measures of HRV, we extracted root mean of the squared successive differences (RMSSD), high frequency power (HF; 0.15-0.4 Hz; absolute values), low frequency power (LF; 0.04-0.15 Hz; absolute values), and ratio of low and high frequency power (LF divided by HF: LF/HF), as well as HR. The HF and LF power were derived by the fast Fourier transform (FFT) spectrum analysis.

.

## 2.4 Data preparation

All physiological measures except HR showed a skewed distribution and were log-transformed to remove this skewness. Two outliers with extreme LF and HF values were removed from all HRV variables.

## 2.5 Statistical analyses

The relations between psychological and physiological measures were investigated by graphical and computational methods. We used scatterplots between all measures and added bidirectionally non-parametrically smoothed curves (Nadaraya-Watson kernel regression (33), Gaussian kernel, bandwidth: $b=0.6SD$). Here, stronger association should result in more similar curves while a weak association should lead to orthogonal regression lines.

As main criterion for the association between the measures we chose the Spearman correlation. The correlations were interpreted as small (0.10), medium (0.30), and large (0.50), according to Cohen's guidelines (34). They were computed both for the entire sample and stratified for the younger and the older participants. Since there is no established age cut-off we first screened by a moving cut-off for age where both groups contained at least 10 participants. The relation between age cut-off and correlations was very difficult to interpret. Therefore, to avoid over-interpretation of the data we decided to use a median cut-off at the age of 28. This seemed to represent the data well and also guaranteed a sufficient size of both groups, so that we base the reported results and the discussion on it. We tested for all correlations if they were significantly different from 0.

The general significance level was set to 0.05. Due to large amount of test we had to take into account multiple effects. The Bonferroni adjustment for all 45 pairwise tests would have been too conservative because the tests were not independent. Thus, we based the adjustment on the number of investigated variables, leading to a marginal significance level of 0.005 (Bonferroni for 10 tests).

Additionally, we estimated linear models with the psychological measures as outcome and the physiological measures as predictor. Here, we computed a fully adjusted model containing all physiological measures as well as unadjusted models for each physiological measure. All models were controlled for age (continuous). Although age was confirmed to be an important confounding variable, keeping age as a continuous confounder in the regression models did not improve the results compared to the correlation analyses. Consequently, we decided to present the analysis stratified by age, as described above. All computation was done by Matlab 7.10 (Mathworks Inc.).

## iii. Results

A descriptive overview of the measured variables is seen in Table 1 and the results of the graphical and the correlation analysis for all participants in Figure 1. Considering the entire sample we can clearly distinguish two blocks of variables. The first block is represented by the lower left square of 5x5 correlations between psychological and physiological measures, those that were of the main interest in this study. These correlations are all between zero and the small range; their low values are also indicated by nearly black color. None of these correlations is significant (not even with the unadjusted significance level 0.05). The other block includes all correlations outside this square, those within measure type. Here, all correlations within the psychological and almost all within the physiological measures are in the medium or large range and significant even at the adjusted significance level. As the only exception, LF/HF was not significantly correlated to HR and LF (correlation values zero to small).

The results of the graphical and the correlation analyses for the younger participants are seen in Figure 2 and for the older ones in Figure 3.

Comparing the psychological with the physiological measures we observed overall a weak pattern for the younger group. Two significant positive correlations of HR with STAI-Trait ($r=0.39$, $p=0.0042$) and HADS-A ($r=0.52$, $p=0.0002$) were in the medium to large range. STAI-State and HADS-D were not significantly correlated with HR but show the same tendencies (with estimates in the medium range). Even if not significant with respect to the adjusted level, we observed weak tendencies for a similar behaviour of the STAI- and the HADS-variables, i.e. negative correlation to RMSSD, LF and HF and positive correlations to LF/HF for the younger group (in the small to medium range). Q-LES-Q had for the younger group a weak tendency to correlate positively to RMSSD , LF and HF (medium range) and not to HR or LF/HF (below the small range). For the older group we could not see tendencies of a correlation between physiological and psychological variables (all in the small range or below). The regression analysis did not lead to different results such that we abstain from reporting the details.

## iv. Discussion

### 4.1 Findings

Overall, this study did not find convincing evidence of a clear general relationship between HR/HRV and depression or other aspects of mental health in adult male prisoners. This was in spite of avoiding some of the main limitations of previous studies: Most importantly, the present study had a larger sample size and therefore higher test power than the individual studies combined in the previous meta-analysis (15). Further innovative aspects of this study were that it used advanced statistical methods that could detect non-linear relations, and that it analyzed several measures of both HRV and mental health. However, it would be premature to conclude that there is no relationship, even in this specific population of male prisoners. The relationship is apparently dependent on different conditions.

First, we observed and partially described an age dependency: In younger participants, HR correlated significantly with trait anxiety and general anxiety; tendencies were also found for state anxiety and depression symptoms. These correlations were positive, suggesting that higher HR indicates higher levels of anxiety and possibly depression symptoms. In contrast to HR, none of the HRV correlated significantly with any of the psychological measures, although some tendencies were observed among younger participants. These tendencies concerned RMSSD, HF, and LF/HF but not LF. These correlations suggest that lower HRV may indicate higher anxiety and depression symptoms as well as poorer quality social relationships in young adults. Thus, for younger adults the results are in line with previous research investigation the relationship between HRV and aspects of mental health (15, 19, 20). However, in older adults, neither HR nor HRV seemed to be a useful indicator of mental health, at least in this study. Generally, “it is likely that any relationship between HR, HRV and mood and anxiety is small to moderate at most” (Andrew Kemp, personal correspondence, July 2014). Figure 4 illustrates the correlations in the present study in direct comparison to previous meta-analytic findings (15) for depression symptoms, the variable studied most extensively. The estimate from the previous meta-analysis (first row in the figure) summarizes the correlation of any HRV measure with depression symptoms. The present study found correlations in similar size for RMSSD, LF, and HF among younger participants, and also for HR (with opposite sign, as hypothesized), but not for LF/HF and not for older participants.

Second, our sample was different from previous samples in that the present sample included a range of severity levels as not all of our participants were at clinical levels (21). However, clear correlations within the two blocks of physiological and psychological measures suggest that our negative findings were not due to limited variation within our sample. This may indicate that the range of the measures has an impact on the relationship between the investigated psychological and HR/HRV-measures.

Third, potential confounding factors that were not available in the present study may have masked existing associations. Numerous confounding variables have been suggested, including but not limited to cardiovascular disease (16, 17), drug abuse (21), medication, physical fitness, body mass index, respiration, restrictions prior to ECG recording, and many others (35). Some of these may be especially relevant in an older cohort (cardiovascular disease, but also hypertension, diabetes, dyslipidemia), perhaps particularly in a prisoner population. In relation to the latter, the present study did not control for antisocial behavior or psychopathy. Prisoners are very often associated with antisocial behavior, and the relationship between antisocial behavior and HRV is complicated and not clearly understood (36). Antisocial behavior might be associated with both low resting HR (22) as well as high level of anxiety (23). However, the relationship between antisocial behavior and HRV reactivity to stress-situations may be different (e.g. increased HR in response to stress compared to controls (37). One reason for these limitations is that the data for this study were not specifically collected with a view to investigating the relationship between measures of mental health and HR/HRV; however, it should be noted that it will always be possible to find potential confounders post hoc (and the zeal to find them may be especially high when findings are negative). Thus, we are not able to give any final answers, but we have presented arguments for the need of a proper study to understand this relationship. Such a study should consider to control for age and possibly other confounding variables.

Consequently, the findings from this study are less positive than those of previous studies. Interestingly, HR appeared to be a better indicator of mental health than HRV, but more clearly so in younger participants. Given that HRV as a more sophisticated measure has attracted much more attention in the research literature than HR, publication bias should be considered as one possible explanation. Bias can occur on two levels: selective publication of positive studies, and selective publication of positive measures within studies. In spite of initiatives to encourage publication of negative findings, reviewers, editors and even authors themselves tend to be more critical of study methods if findings were negative (38).

### 4.2 Technical remarks

One issue of studies with multiple dependent tests is the selection of the significance level. In Appendix A.1 we report the implications of our chosen marginal level (0.005) on the familywise error rates. Conversely, an adjustment of the significance level always leads to a reduction of power. In Appendix A.2 we show that we had sufficient power using the adjusted level.

### 4.3 Implications for research and use of biomarkers

Like other potential biomarkers for mental health (3), much basic research remains to be done before application in clinical trials can seriously commence. Even as a surrogate outcome, the relation of the biomarker candidate with the clinical outcome needs to be confirmed with greater certainty than is presently the case. There are numerous hypotheses about the role of HRV in describing mental health (12, 15, 19, 20), which are interesting, but our findings suggest, in line with others (35), that caution is needed when using HRV across a range of participants with different characteristics. There seems to be no simple relation between HR/HRV and specific aspects of mental health; rather, the relation between them may be unspecific and complicated by a large number of confounders (of which our study has confirmed one). In addition, the issue of publication bias may deserve increased attention also in correlational studies of biomarkers. Prospective registration of studies, which has become a requirement in clinical trials (18), may also serve to improve trust in research findings in this area (39).

Due to the remaining uncertainties around the validity of HRV as a biomarker of mental health across populations and age groups, clinical trials of intervention effects should continue to use traditional clinical measures, such as questionnaires and clinical assessments, as primary outcomes (3). Physiological measures of mental health can serve as additional surrogate outcomes but their correlation with clinical outcomes should be carefully examined. More generally, the link between physical and mental health remains a challenge. As new potential biomarkers continue to be proposed and developed, these will again create a need for thorough and critical evaluation of their actual value in clinical applications. Future research might bring more satisfying answers to the ability of HRV and other physiological measures to indicate aspects of mental health.

**Acknowledgements**
The study was funded by a grant from the Norwegian Directorate for Health and Social Affairs and GC Rieber Foundation. Intramural funding was provided by Bjørgvin prison, the University of Bergen, and Uni Research. Leif Waage, Harald Åsaune, and Brynjulf Stige helped to establish the study; Anita Lill Hansen with HRV technology; Liv Gunnhild Qvale and Fiona Kirkwood Brown with data collection. Andrew Kemp and Carmilla Licht provided valuable comments on an earlier version of this paper.

**Declaration of Interest**
The authors reported no biomedical financial interests or potential conflicts of interest.

**References**

1. Gold C, Rolvsjord R, Mössler K, Stige B. Reliability and validity of a scale to measure interest in music among clients in mental health care. Psychol Music. 2013 Sep;41(5):665-82. PubMed PMID: WOS:000323739600009. English.
2. Luecken LJ, Gallo LC, editors. Handbook of physiological research methods in health psychology. Thousand Oaks, CA: Sage; 2008.
3. Gold C, Fachner J, Erkkilä J. Validity and reliability of electroencephalographic frontal alpha asymmetry and frontal midline theta as biomarkers for depression. Scand J Psychology. 2013;54(2):118-26.
4. Fachner J, Gold C, Erkkilä J. Music therapy modulates fronto-temporal activity in the rest-EEG in depressed clients. Brain Topogr. 2013 Apr;26(2):338-54. PubMed PMID: 22983820.
5. Haslbeck FB. Music therapy for premature infants and their parents: an integrative review. Nord J Music Ther. 2012;21(3):203-26.
6. Teckenberg-Jansson P, Huotilainen M, Pölkki T, Lipsanen J, Järvenpää AL. Rapid effects of neonatal music therapy combined with kangaroo care on prematurely-born infants. Nord J Music Ther. 2011;20(1):22-42. PubMed PMID: WOS:000287080000003. English.
7. Gold C. Mental health assessments and multicultural encounters: Hearing the grass grow. Nord J Music Ther. 2013;22(2):91-2. PubMed PMID: WOS:000320510200001. English.
8. O'Kelly J, Magee WL. Music therapy with disorders of consciousness and neuroscience: the need for dialogue. Nord J Music Ther. 2013;22(2):93-106.
9. Clark I, Harding K. Psychosocial outcomes of active singing interventions for therapeutic purposes: a systematic review of the literature. Nord J Music Ther. 2012;21(1):80-98.
10. Miller EB, Goss CF. Trends in physiological metrics during Native American flute playing. Nord J Music Ther. 2014.
11. Warth M, Kessler J, Koenig J, Hillecke TK, Wormit AF, Bardenheuer HJ. Methodological challenges for music therapy controlled clinical trials in palliative care. Nord J Music Ther. 2015.
12. Thayer JF, Ahs F, Fredrikson M, Sollers JJ, Wager TD. A meta-analysis of heart rate variability and neuroimaging studies: Implications for heart rate variability as a marker of stress and health. Neurosci Biobehav R. 2012 Feb;36(2):747-56. PubMed PMID: WOS:000300747600002. English.
13. Aberg MAI, Nyberg J, Toren K, Sorberg A, Kuhn HG, Waern M. Cardiovascular fitness in early adulthood and future suicidal behaviour in men followed for up to 42 years. Psychol Med. 2014 Mar;44(4):779-88. PubMed PMID: WOS:000332954700011. English.
14. Lemogne C, Thomas F, Consoli SM, Pannier B, Jego B, Danchin N. Heart Rate and Completed Suicide: Evidence From the IPC Cohort Study. Psychosom Med. 2011 Nov-Dec;73(9):731-6. PubMed PMID: WOS:000297205700002. English.
15. Kemp AH, Quintana DS, Gray MA, Felmingham KL, Brown K, Gatt JM. Impact of depression and antidepressant treatment on heart rate variability: A review and meta-analysis. Biol Psychiatr. 2010;67:1067-74.
16. Licht CMM, Pennix BW, deGeus EJC. To include or not to include? A response to the meta-analysis of heart rate variability and depression. Biol Psychiatr. 2011;69:e1.
17. Kemp AH, Quintana DS, Gray MA. Is heart rate variability reduced in depression without cardiovascular disease? Biol Psychiatr. 2011;69:e3-e4.
18. Higgins JPT, Green S, editors. Cochrane handbook for systematic reviews of interventions. Chichester, UK: Wiley-Blackwell; 2008.

19. Chalmers JA, Quintana DS, Abbott MJA, Kemp AH. Anxiety disorders are associated with reduced heart rate variability: a meta-analysis. Front Psychiatry. 2014;5(80).
20. Schwerdtfeger A, Friedrich-Mai P. Social Interaction Moderates the Relationship Between Depressive Mood and Heart Rate Variability: Evidence From an Ambulatory Monitoring Study. Health Psychol. 2009 Jul;28(4):501-9. PubMed PMID: WOS:000267971100015. English.
21. Gold C, Assmus J, Hjørnevik K, Qvale LG, Brown FK, Hansen AL, et al. Music therapy for prisoners: Pilot randomised controlled trial and implications for evaluating psychosocial interventions. Int J Offender Ther Comp Criminol. 2014;58(12):1520-39.
22. Raine A. Biosocial studies of antisocial and violent behavior in children and adults: A review. J Abnorm Child Psych. 2002 Aug;30(4):311-26. PubMed PMID: WOS:000175923600001. English.
23. Hansen AL, Stokkeland L, Johnsen BH, Pallesen S, Waage L. The relationship between the psychopathy checklist-revised and the MMPI-2: A pilot study. 2013;112(2):445-57.
24. Fazel S, Danesh J. Serious mental disorder in 23 000 prisoners: a systematic review of 62 surveys. Lancet. 2002;359(9306):545-50.
25. Priebe S, Frottier P, Gaddini A, Kilian R, Lauber C, Martínez-Leal R, et al. Mental health care institutions in nine European countries, 2002 to 2006. Psychiat Serv. 2008;59(5):570-3.
26. Olssøn I, Mykletun A, Dahl AA. The hospital anxiety and depression rating scale: A cross-sectional study of psychometrics and case finding abilities in general practice. BMC Psychiatry. 2005;5(46).
27. Spielberger C, Gorsuch R, Lushene R. Manual for the State-Trait Anxiety Inventory. Palo Alto, CA: Consulting Psychologists Press; 1970.
28. Barnes LLB, Harp D, Jung WS. Reliability generalization of scores on the Spielberger State-Trait Anxiety Inventory. Educ Psychol Meas. 2002;62(4):603-18.
29. Zigmond AS, Snaith RP. The Hospital Anxiety and Depression Scale. Acta Psychiatr Scand. 1983;67(6):361-70.
30. Bjelland I, Dahl AA, Haug TT, Neckelmann D. The validity of the Hospital Anxiety and Depression Scale: An updated literature review. J Psychosom Res. 2002;52(2):69-77.
31. Endicott J, Nee J, Harrison W, Blumenthal R. Quality of Life Enjoyment and Satisfaction Questionnaire: a new measure. Psychopharmacol Bull. 1993;29(2):321-6.
32. Brage S, Brage N, Franks P, Ekelund U, Wareham NJ. Reliability and validity of the combined heart rate and movement sensor Actiheart. Eur J Clin Nutr. 2005;59:561-70.
33. Hastie T, Tibsharani R, Friedman J. The Elements of Statistical Learning. New York, NY: Springer; 2001. 191 p.
34. Cohen J. Statistical power analysis for the behavioral sciences. 2nd ed. Hillsdale, NJ: Lawrence Erlbaum; 1988.
35. Quintana DS, Heathers JAJ. Considerations in the assessment of heart rate variability in biobehavioral research. Front Psychol. 2014 Jul 22;5. PubMed PMID: WOS:000339396300001. English.
36. Hansen AL, Johnsen BH, Thornton D, Waage L, Thayer JF. Facets of psychopathy, heart rate variability and cognitive function. J Pers Disord. 2007 Oct;21(5):568-82. PubMed PMID: WOS:000253432200008. English.
37. Patrick CJ, Bernat EM. From markers to mechanisms: Using psychopysiological measures to elucidate basic processes underlying aggressive externalizing behavior. In: Hodgins S, Viding E, Plodowski A, editors. The neurobiological basis of violence: Science and rehabilitation. New York, NY: Oxford University Press; 2009. p. 223-50.

38. Emerson GB, Warme WJ, Wolf FM, Heckman JD, Brand RA, Leopold SS. Testing for the Presence of Positive-Outcome Bias in Peer Review A Randomized Controlled Trial. Arch Intern Med. 2010 Nov 22;170(21):1934-9. PubMed PMID: WOS:000284480000015. English.
39. Cochrane Diagnostic Test Accuracy Working Group. Handbook for DTA Reviews. http://srdta.cochrane.org/handbook-dta-reviews (last accessed 19 July, 2013)2009 19 July, 2013]. Available from: http://srdta.cochrane.org/handbook-dta-reviews.

**Tables**

**Table 1. Sample characteristics**

| Variable | N | Mean (SD) | Younger | Older |
|---|---|---|---|---|
| Age (years) | 113 | 31.38 (10.72) | 22.89 (2.57 | 39.43 (9.17) |
| Expected stay (days) [1] | 100 | 42 [4, 840] | 30.5 [7, 472] | 46.5 [4, 840] |
| State anxiety (STAI-State) | 110 | 38.84 (11.91) | 39.64 (12.01) | 38.09 (11.88) |
| Trait anxiety (STAI-Trait) | 109 | 43.23 (10.09) | 45.51 (9.61) | 41.07 (10.14) |
| Anxiety (HADS-A) | 106 | 7.59 (4.36) | 7.88 (3.79) | 7.33 (4.84) |
| Depression (HADS-D) | 106 | 5.51 (3.72) | 5.60 (3.24) | 5.43 (4.16) |
| Social relationship (Q-LES-Q) | 109 | 40.83 (6.69) | 40.77 (6.27) | 40.88 (7.10) |
| Anxiety above cut-off (HADS-A≥8) [2,3] | 106 | 48 (45%) | 27 (52.9%) | 21 (38.2%) |
| Depression above cut-off (HADS-D≥8) [2,3] | 106 | 31 (29%) | 18 (34.6%) | 13 (24.1%) |
| HR [5] | 108 | 77.44 (12.57) | 77.00 (13.52) | 77.84 (11.72) |
| RMSSD (log) [4,5] | 109 | 3.63 (0.71) | 3.80 (0.65) | 3.48 (0.73) |
| LF (log) [4,5] | 109 | 6.83 (1.17) | 7.07 (1.02) | 6.61 (1.26) |
| HF (log) [4,5] | 109 | 5.87 (1.47) | 6.32 (1.38) | 5.46 (1.45) |
| LF/HF (log) [4,5] | 109 | 1.00 (0.83) | 0.72 (0.81) | 1.27 (0.77) |

*Note.* [1] Median [range]. [2] Count (percent). [3] Cutoff for HADS-A and HADS-D based on Olssøn et al (2005). [4] Natural logarithm. Geometric means in the original scale were: RMSSD 39.0, LF 922.9, HF 354.1, LF/HF 2.73. [5] Two outliers with extreme values removed.

# Figures

## Figure 1. All participants: Matrix plot of the relation between psychological and physiological measurements.

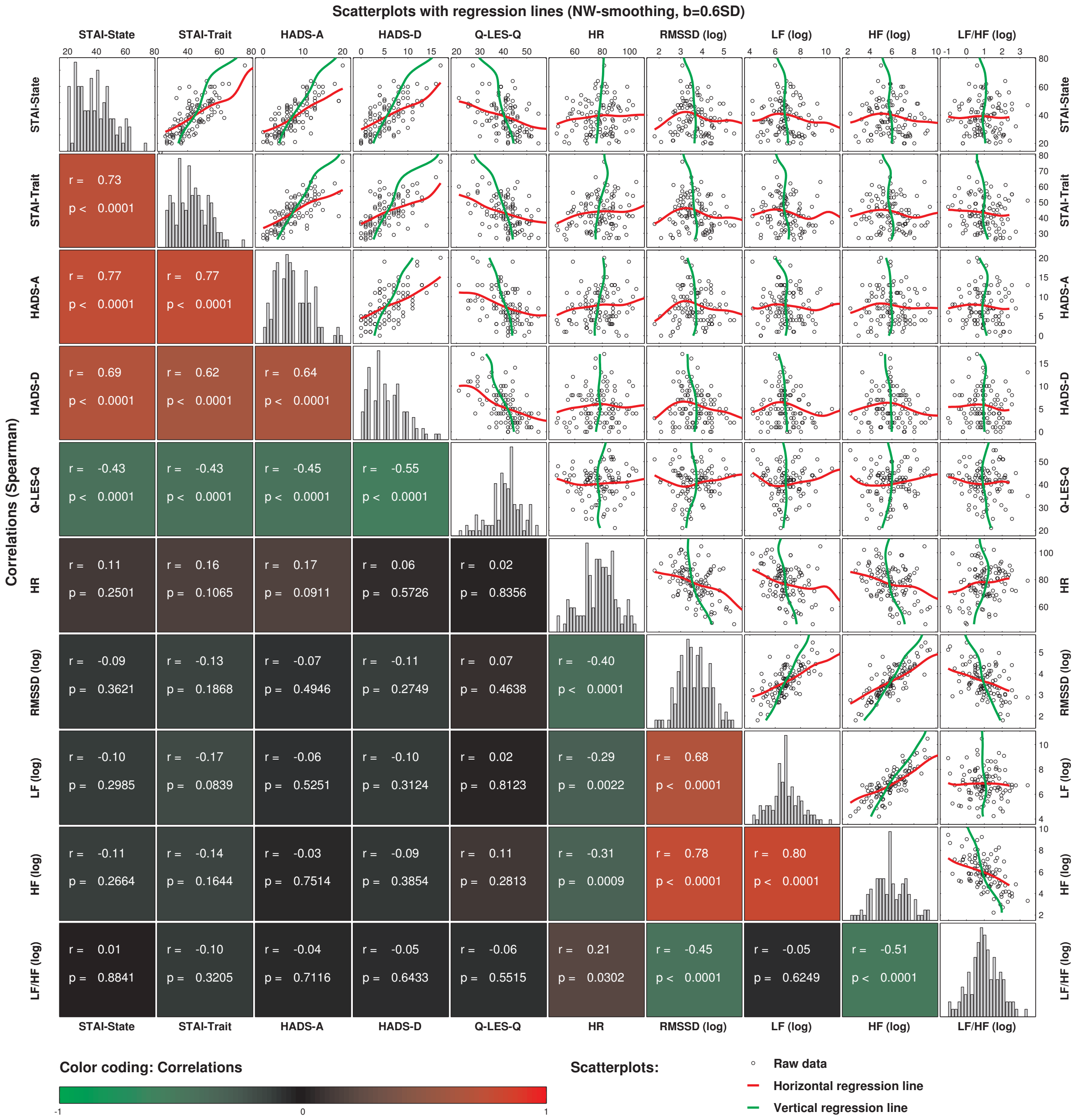

The elements of the matrix plot are: (i) Subplots in the diagonal: Histograms of each variable. (ii) Upper right triangle: Bivariate scatter plots with regression lines (Nadaraya-Watson kernel regression, Gaussian kernel, bandwidth: *b=0.6SD*). The red lines are the regression line for the y-axis depending on the x-axis; the green lines are regression lines for the x-axis depending on the y-axis. (iii) Lower left triangle: Spearman correlations between variables as numbers and coded as colours with the corresponding p-values for the test if they equal 0.

**Figure 2. Younger participants (<28years): Matrix plot of the relation between psychological and physiological measurements.**

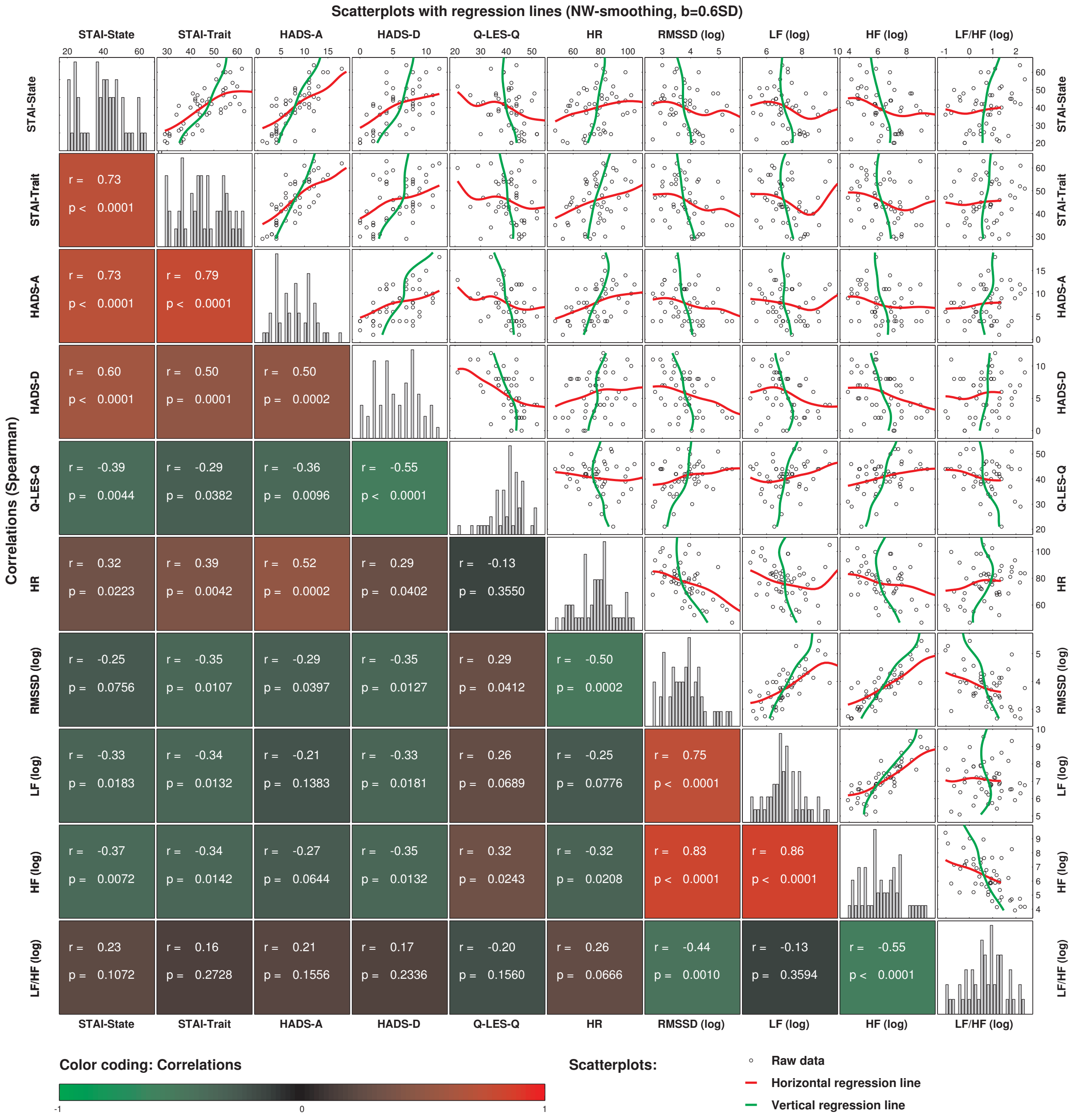

The elements of the matrix plot are explained in Figure 1.

**Figure 3. Older participants (≥28years): Matrix plot of the relation between psychological and physiological measurements.**

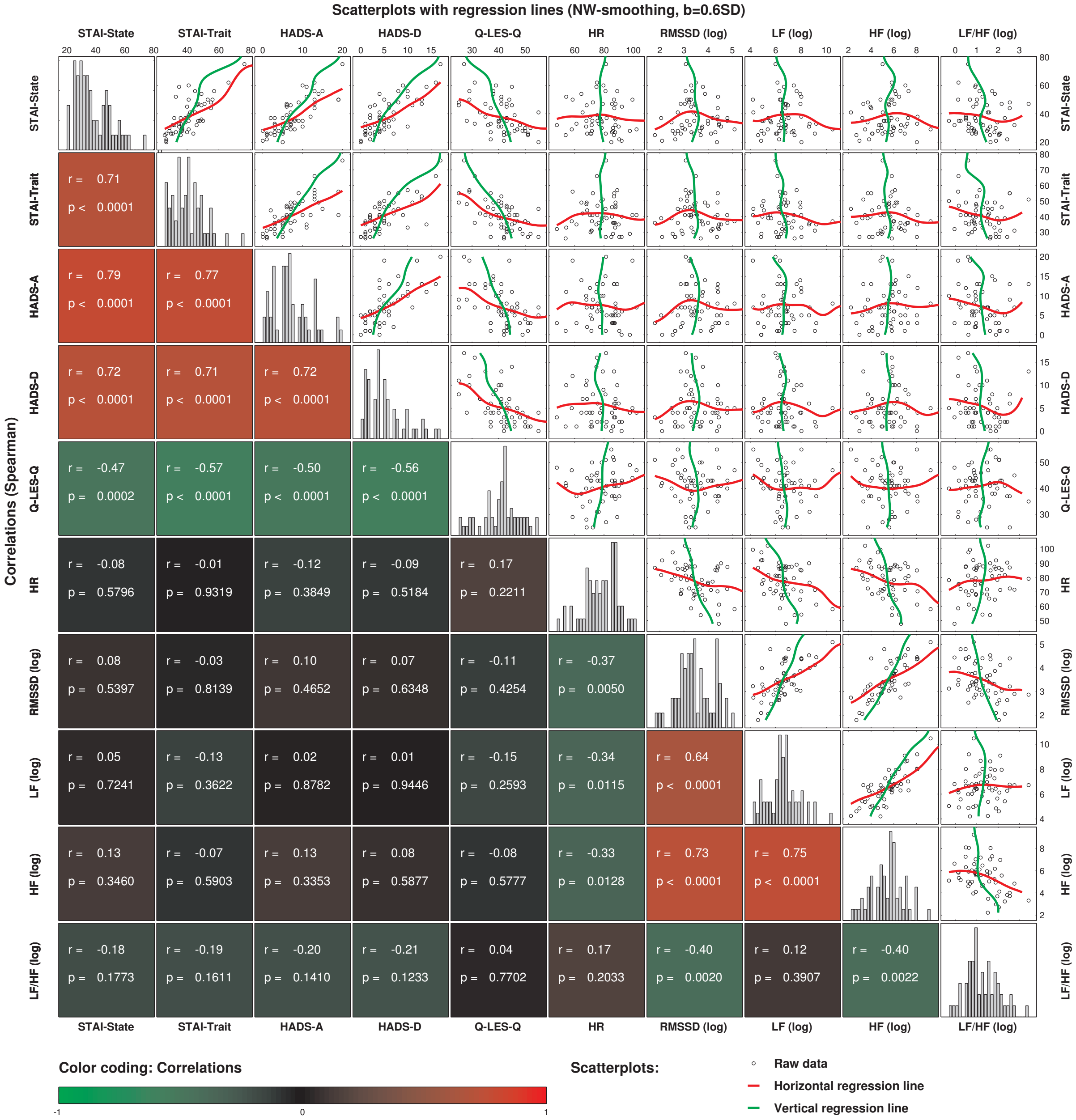

The elements of the matrix plot are explained in Figure 1.

**Figure 4. Comparison of this study's findings to a previous meta-analysis**

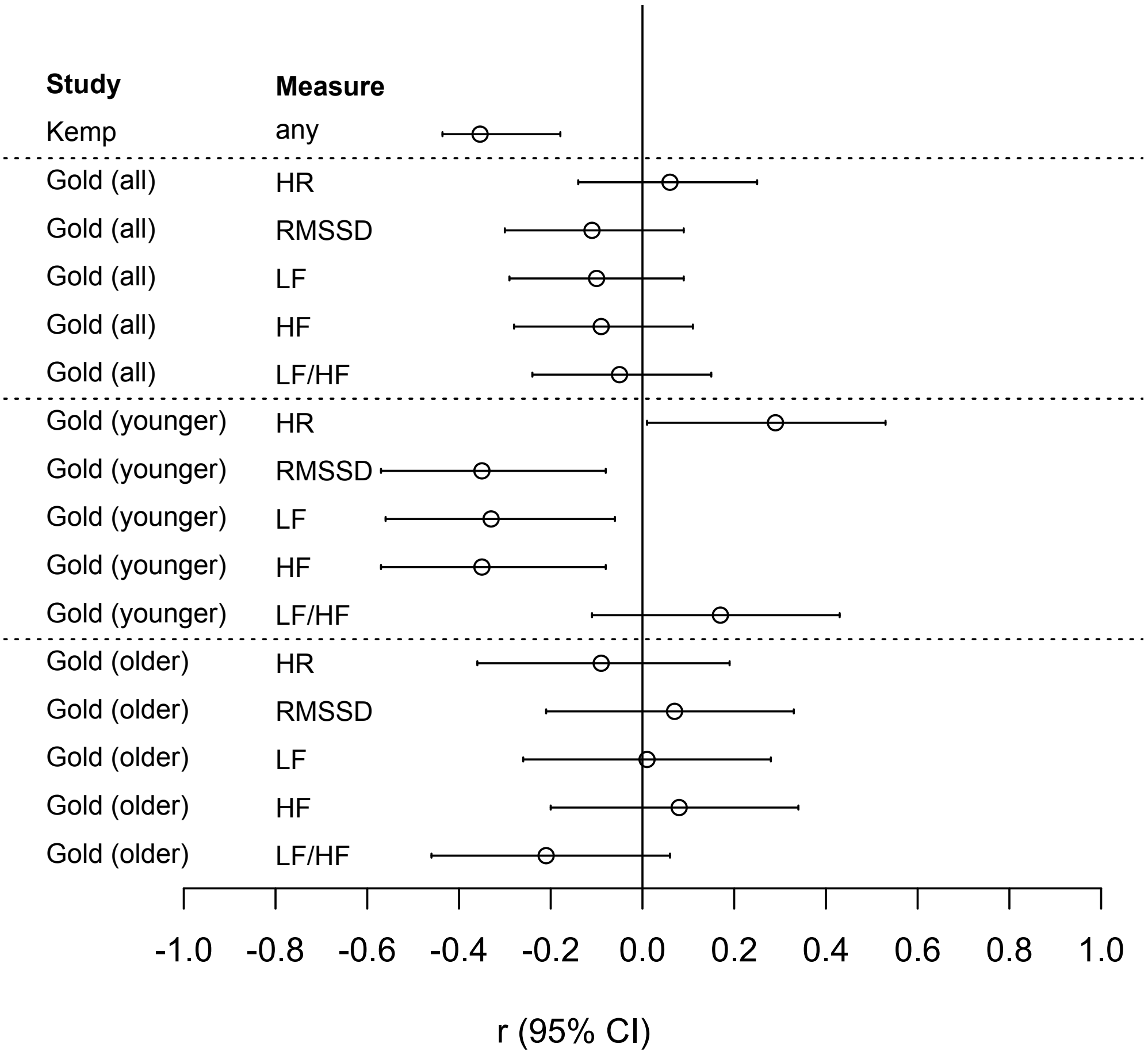

Showing correlations with 95% confidence intervals. Kemp – previous meta-analysis (15); Gold – this study; younger – participants <28 years; older – participants ≥28 years.

# A. Appendix

## A.1 Determination of the marginal significance level

If one does a number of tests, the probability to obtain a type I error in at least one of them increases with each additional test (using the same significance level). This probability is called familywise error rate (FWER). With a level of 0.05 and independent tests, one can expect that about 5% of the tests will be wrongly rejected. One approach to handle that problem is to adjust the marginal significance level (level for each single test) such that the FWER is equal or less than 0.05. The most common approach to do that is the Bonferroni adjustment. This adjustment is an approximation, i.e. it drops below the 0.05-limit of FWER – the more tests the lower is the FWER, e.g. for 10 independent test the Bonferroni-adjustment leads to a FWER of 0.049. Thus, the Bonferroni adjustment is very conservative and becomes more conservative with increasing number of tests.

Dependencies between the tests reduce the multiple effects, i.e. the Bonferroni adjustments are even more conservative. In this case it is often difficult to describe the dependencies to determine the correct appearance of multiple effects. We did a simulation study where we tested the pairwise correlations of 10 independent variables (i.e. under null hypothesis) with 10,000 repetitions for 100 participants. We computed the relation between marginal significance level and the FWER and obtained the following results:

| Marginal level | FWER | Type |
|---|---|---|
| 0.0011 | 0.025 | Bonferroni for 45 tests |
| 0.0024 | 0.050 | Optimal level for FWER=0.05 |
| 0.0050 | 0.100 | Level used in this study |

These results show that the Bonferroni level is far too conservative while the optimal level is lower than the one we used. The level we used leads to FWER of 0.1 – this matches exactly the 0.1 level if one assumes one-sided tests (adjusted for correct multiple effects). One-sided tests may be justified because previous studies did indicate likely directions of the correlations we examined.

This simulation study applies to independent variables. However, there were correlated variables in the data, e.g. between the psychological measures, which reduced the effects of multiple testing. That is, the marginal level should be higher than the optimal level for independent variables (0.0024).

Overall, the results of this simulation show that the marginal level used in this study provides a reasonable compromise between possible over-adjustment and under-adjustment.

## A.2 Power of the tests

With the sample size of 113, the marginal level of 0.005 translates for a single test to 80% power for correlations of at least 0.34. Using the nominal 5% significance level, the same test power of 80% would be reached for correlations of at least 0.26. The loss of power due to the adjustment for multiplicity was therefore relatively modest. Test power was sufficient to detect medium or larger correlations.